\documentclass [11pt] {article}
\usepackage[utf8]{inputenc}
\usepackage[T1]{fontenc}
\usepackage{xcolor,amsmath,amssymb}
\usepackage{geometry}
\usepackage{graphicx}
\usepackage{enumerate}

\definecolor{darkblue}{rgb}{0,0,0.6}
\definecolor{darkred}{rgb}{0.6,0,0}
\usepackage[colorlinks=true,urlcolor=darkblue,citecolor=darkblue,linkcolor=darkred,hyperfootnotes=false]{hyperref}

\usepackage{authblk}

\usepackage{cmbright}

\renewcommand{\boldsymbol}[1]{\mathbold{#1}}
\renewcommand{\mathrm}[1]{\textrm{\sf #1}}

\newcommand{\ind}[1]{_{\mathrm{#1}}}

\newcommand{\ff}{\boldsymbol{f}}
\newcommand{\FF}{\boldsymbol{F}}
\newcommand{\jj}{\boldsymbol{j}}
\newcommand{\kk}{\boldsymbol{k}}
\newcommand{\qq}{\boldsymbol{q}}

\newcommand{\mcV}{\mathcal{V}}
\newcommand{\vv}{\boldsymbol{v}}
\newcommand{\xx}{\boldsymbol{x}}
\newcommand{\yy}{\boldsymbol{y}}

\newcommand{\eeta}{\boldsymbol{\eta}}

\newcommand{\ppsi}{\boldsymbol{\psi}}

\newcommand{\Pe}{\mathrm{Pe}}

\newcommand{\transp}{^\mathrm{T}}

\title{Mean-field microrheology of a very soft colloidal suspension: inertia induces shear-thickening}

\author{Vincent D\' emery}
\affil{Department of Physics, University of Massachusetts, Amherst, MA 01003, USA}
\affil{Laboratoire de Physico-Chimie Théorique, UMR CNRS Gulliver 7083, ESPCI, Paris, France}

\begin{document}

\maketitle

\begin{abstract}
Colloidal suspensions have a rich rheology and can exhibit shear-thinning as well as shear-thickening.
Numerical simulations recently suggested that shear-thickening may be attributed to the inertia of the colloids, besides the hydrodynamic interactions between them.
Here, we consider the ideal limit of a dense bath of soft colloids following an underdamped Langevin dynamics.
We use a mean-field equation for the colloidal density to get an analytical expression of the drag force felt by a probe pulled at constant velocity through the suspension.
Our results show that inertia can indeed induce shear-thickening by allowing density waves to propagate through the suspension.
\end{abstract}

\section{Introduction}\label{}

Suspensions of colloids or droplets have a rich rheology~\cite{Mewis2011,Coussot2005}.
Whereas they can have a yield stress at volume fractions above the glass or jamming transitions~\cite{Ikeda2012}, they behave as simple Newtonian fluids at moderate densities and small shear rates.
Upon increasing the shear rate, their viscosity can then decrease (shear-thinning) or increase (shear-thickening)~\cite{Laun1984,Wagner2009}.

\begin{figure}
\begin{center}
\includegraphics[width=.45\linewidth]{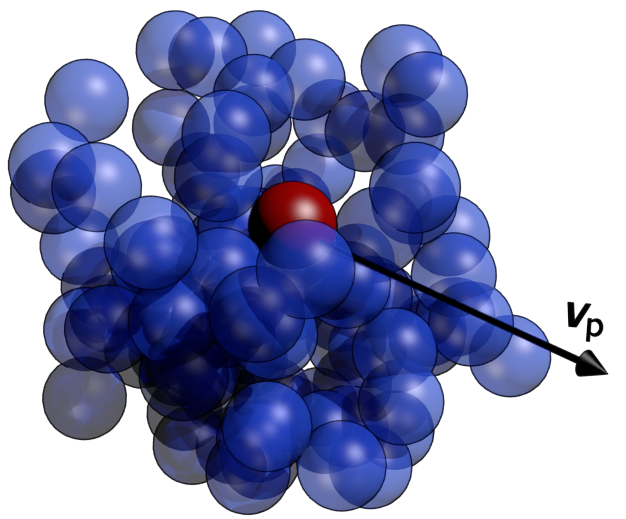}
\end{center}
\caption{(color online)  Illustration of the system studied: a probe (red) is pulled at constant velocity through a dense bath of soft colloids with inertia.}
\label{fig:illustration}
\end{figure}

Experimentally, the suspension rheology can be investigated using macrorheology or microrheology. 
In macrorheology, a global shear rate $\dot\gamma$ is applied, and the resulting shear stress $\tau$ is measured ; the viscosity is then defined as $\eta=\tau/\dot\gamma$~\cite{Bagnold1954,Boyer2011b}.
In microrheology, the motion of a small probe in the medium is observed~\cite{Waigh2005,Squires2010}. 
Notably, in active microrheology, the probe is placed in an optical or magnetic trap and pulled at constant velocity $v$ (or at constant force $F$) through the medium~\cite{Habdas2004,Wilson2011b,Meyer2006,Sriram2010,Puertas2014}~(see Fig.~\ref{fig:illustration}). Measuring the drag force $F$ on the probe (or its average velocity $v$) and the corresponding drag coefficient $\lambda=F/v$, one can use the Stokes formula to deduce the viscosity: $\lambda=6\pi\eta a\ind{p}$, where $a\ind{p}$ is the radius of the probe.
These two approaches have their theoretical counterpart both in numerical simulations (\cite{Ikeda2012,Kawasaki2014,Trulsson2014,Trulsson2012} for macrorheology, \cite{Carpen2005,Winter2012} for microrheology) and analytical computations (\cite{Bagnold1954,Fuchs2002} for macrorheology, \cite{Squires2005,Khair2006} for microrheology).

Shear-thinning, which is ubiquitous in experiments, is also present in most of the analytical computations both for dilute~\cite{Squires2005,Khair2006} and dense suspensions~\cite{Fuchs2002,Harrer2012,Gazuz2009,Gnann2011,Gazuz2013,Demery2014c} (see~\cite{Puertas2014} for a review). It is commonly associated with the disruption of the equilibrium microscopic structure, which gives the solution a large viscosity.
On the other hand, various forms of shear-thickening exist and they are difficult to describe theoretically~\cite{Wagner2009,Brown2014}. 
While discontinuous shear-thickening may arise due to a dynamic jamming transition~\cite{Brown2009,Wyart2014,Fall2015}, a softer, continuous shear-thickening is induced by the formation of hydro-clusters due to the lubrication forces, which hold the particles together~\cite{Melrose2004}. 
However, this mechanism has a negligible effect on soft-particles~\cite{Bergenholtz2002,Khair2006,Wagner2009}: grafting polymer brushes to hard colloids can considerably delay shear-thickening~\cite{Mewis2001}.

A recent numerical work addressed the role of colloids inertia on the suspension rheology, neglecting the hydrodynamic interactions, and showed that it can induce shear-thickening as long as the system is sufficiently far from jamming~\cite{Kawasaki2014}.
In this article, we provide an analytical derivation of this effect in the limit of very dense and soft colloids.
We use microrheology to investigate the properties of this medium: we compute the drag force felt by a probe pulled at constant velocity through the suspension.
We obtain an analytical expression for the drag coefficient $\lambda$, which displays shear-thickening induced by inertia.

This article is organized as follows.
The model of underdamped Langevin colloids is presented in Sec.~\ref{sec:model}.
A linearized equation for the coarse-grained density field of the bath is obtained in Sec.~\ref{sec:deriv_inertial_dean}.
This linearized equation is used to compute the response of the bath to the passage of the probe at constant velocity in Sec.~\ref{sec:response_to_probe}.
The stationnary bath density around the probe is computed (Subsec.~\ref{sub:bath_density}), from which the drag felt by the probe is deduced (Subsec.~\ref{sub:drag}).
The analytical expressions are computed numerically and discussed in Subsec.~\ref{sub:numerical_computation}.
Our results are discussed in terms of the different timescales involved in the problem in Subsec.~\ref{sub:timescales}.
Finally, the effect of the mode of driving is mentioned in Subsec.~\ref{sub:cv_cf}.


\section{Model}\label{sec:model}

We consider $N$ colloids in a $d$-dimensional bath with positions $\xx_i(t)$ interacting via the pair potential $V(\xx)$ and evolving according to Langevin dynamics, that we write in the form
\begin{equation}\label{eq:langevin}
m\ddot \xx_i(t)+\lambda\ind{s}\dot \xx_i(t)=\FF_i(t)+\eeta_i(t),
\end{equation}
where $m$ is the mass of the colloids and $\lambda\ind{s}$ the friction coefficient with the solvent. We can thus define the damping time associated to the inertia of the colloids as 
\begin{equation}
\tau=\frac{m}{\lambda\ind{s}}.
\end{equation}
$\ff_i(t)$ is the force on the particle $i$ and $\eeta_i(t)$ is the Gaussian white noise on the particle $i$. The noise is completely defined by its correlation function
\begin{equation}
\left\langle \eeta_i(t)\eeta_j(t')\transp \right\rangle=2T\lambda\ind{s}\delta_{ij}\delta(t-t')\mathbf{1},
\end{equation}
where $T$ is the temperature (the Boltzmann constant is set to $k\ind{B}=1$).
The force is given by the gradient of the potential created by the other colloids,
\begin{equation}
\FF_i(t)=-\sum_j\nabla_i V(\xx_i(t)-\xx_j(t)).
\end{equation}
The volume $\mathcal{V}$ of the box containing the colloids is taken to infinity, keeping the density $\rho_0=N/\mathcal{V}$ constant.

A probe, which interacts with the bath colloids with the potential $V\ind{p}(\xx)$, is pulled at constant velocity $\vv\ind{p}$ through the suspension.
The mode of driving, constant velocity instead of constant force, is chosen because it leads to easier analytical computations. 
The average velocity induced by a constant force applied on the probe has been computed without inertia for the bath particles in~\cite{Demery2014c}.
Altough it has been shown that in dilute systems without hydrodynamic interactions the viscosity measured by imposing the velocity is twice the one measured by imposing the force~\cite{Swan2013}, there is no general relation between the results obtained through these two modes of driving~\cite{Puertas2014}.

Without loss of generality, we can set the size of the bath colloids $a$, the thermal energy $T$ and the friction coefficient $\lambda\ind{s}$ to 1 (we keep the temperature explicitly in our expressions to make the role of the temperature clear).
The dimensional counterpart of the dimensionless quantities computed below is obtained by multiplicating them by the appropriate factor of the size of the probe $a$, the thermal energy $k\ind{B}T$ and the friction coefficient of a colloid in the solvent $\lambda\ind{s}$.

\section{Linearized equation for the bath density}\label{sec:deriv_inertial_dean}

Given the colloids positions $\xx_i(t)$ and velocities $\vv_i(t)$, the bath density $\hat\rho(\xx,t)$ and current $\hat\jj(\xx,t)$ are defined by
\begin{align}
\hat \rho(\xx,t) & = \sum_i\delta(\xx-\xx_i(t)),\\
\hat \jj(\xx,t) & = \sum_i\vv_i \delta(\xx-\xx_i(t)).
\end{align}
For colloids with Langevin dynamics~(\ref{eq:langevin}), Nakamura and Yoshimori derived the exact equations satisfied by these two fields~\cite{Nakamura2009}:
\begin{align}
\partial_t\hat \rho & = -\nabla\cdot \hat \jj, \label{eq:nakamura1}\\
\partial_t\hat \jj  & = -\frac{1}{\tau}\hat \jj - \frac{1}{\tau}\hat \rho\nabla(V*\hat \rho)-\nabla\cdot \left(\frac{\hat \jj\hat \jj\transp}{\hat \rho} \right)+\frac{\sqrt{T\hat \rho}}{\tau}\eeta,\label{eq:nakamura2}
\end{align}
where $*$ is the convolution product, $V*\hat\rho(\xx,t)=\int V(\xx-\xx')\hat\rho(\xx',t)d\xx'$, and $\eeta(\xx,t)$ is a Gaussian white noise with correlation function
\begin{equation}
\left\langle \eeta(\xx,t)\eeta(\xx',t')\transp \right\rangle = 2T\delta(\xx-\xx')\delta(t-t')\mathbf{1}.
\end{equation}

Eqs.~\ref{eq:nakamura1} and \ref{eq:nakamura2} are non-linear and contain multiplicative noise, which makes their analytical treatment difficult. This situation is inextricable for the real fields, but the equations can be linearized if one works with the coarse-grained fields $\rho(\xx,t)$ and $\jj(\xx,t)$, which obey~\cite{Das2013}
\begin{align}
\partial_t \rho & = -\nabla\cdot  \jj,\label{eq:rho_cg}\\
\partial_t \jj & = -\frac{1}{\tau} \left[\jj+T\nabla \rho+\rho\nabla(V*\rho)-\sqrt{T\rho}\eeta \right]-\nabla \cdot \left(\frac{\jj \jj\transp}{\rho} \right).\label{eq:j_cg}
\end{align}
In Eq.~\ref{eq:j_cg}, we used the random phase approximation~\cite{Hansen2006} to write the direct correlation function $c(r)$ with the interparticle potential, $c(r)=-V(r)/T$.
The random phase approximation is justified if the pair potential is weak, which is the case that we consider.

Following~\cite{Demery2014c}, we can now linearize Eqs.~\ref{eq:rho_cg} and \ref{eq:j_cg} around a large homogeneous density $\rho_0$: writing
\begin{align}
\rho(\xx,t) & = \rho_0+\rho_0^{1/2}\phi(\xx,t),\\
\jj(\xx,t) & = \rho_0^{1/2}\ppsi(\xx,t)
\end{align}
and taking the limit $\rho_0\to\infty$ with $\rho_0 V(\xx)\to\mcV(\xx)$, we get
\begin{align}
\partial_t\phi  & = -\nabla\cdot\ppsi,\label{eq:lin_rho}\\
\partial_t\ppsi & = -\frac{1}{\tau}\left[\ppsi +T\nabla\phi+\nabla(\mcV*\phi)-\sqrt{T}\eeta\right].\label{eq:lin_j}
\end{align}
A closed second order equation can be obtained for the density field,
\begin{equation}\label{eq:linearized_id}
\tau\partial^2_t\phi+\partial_t\phi = T\nabla^2\phi+\nabla^2(\mcV*\phi)+\sqrt{T}\nabla\cdot\eeta.
\end{equation}
The contribution of inertia in this expression is remarkably simple: it enters only in the first term on the l.h.s. 
Without inertia, this equation reduces to the linearized Dean equation~\cite{Demery2014c}.

\section{Application to a probe pulled at constant velocity}\label{sec:response_to_probe}

\subsection{Bath density around the probe}\label{sub:bath_density}

In order to assess the rheological properties of the suspension, instead of shearing the material globally as in \cite{Kawasaki2014}, we pull a probe particle at constant velocity $\vv\ind{p}$ through the medium, as in \cite{Demery2010}. 


The interaction between the probe and a particle of the bath is given by the potential $V\ind{p}(\xx)=\mcV\ind{p}(\xx)/\rho_0$.
The effect of the probe on the density field can be incorporated in the linearized equation \ref{eq:linearized_id} as in \cite{Demery2014c},
\begin{equation}
\tau\partial_t^2\phi(\xx,t)+\partial_t\phi(\xx,t) 
= T\nabla^2\phi(\xx,t)+\nabla^2 \left[(\mcV*\phi)(\xx,t) \right]+\nabla\cdot\bar\eeta(\xx,t) 
 + \rho_0^{-1/2}\nabla^2[\mcV\ind{p}(\xx-\xx\ind{p}(t)],
\end{equation}
where $\xx\ind{p}(t)=\vv\ind{p}t$ is the position of the probe. Conversely, the force exerted by the bath particles on the probe is
\begin{equation}
\FF(t)=-\rho_0^{-1/2}\nabla \left[(\mcV\ind{p}*\phi)(\xx\ind{p}(t),t) \right].
\end{equation}

We are interested in the average stationnary solution for the field in the reference frame of the particle,
\begin{equation}
\phi^*(\xx)=\langle \phi(\xx+\vv\ind{p}t,t) \rangle;
\end{equation}
it satisfies
\begin{equation}
\tau(\vv\ind{p}\cdot\nabla)^2\phi^*(\xx)-\vv\ind{p}\cdot\nabla\phi^*(\xx)=T\nabla^2\phi^*(\xx)+\nabla^2(\mcV*\phi^*(\xx))+\rho_0^{-1/2}\nabla^2\mcV\ind{p}(\xx).
\end{equation}
In Fourier space, this equation reads
\begin{equation}
\left[\kk^2(T+\tilde\mcV(\kk))-i\vv\ind{p}\cdot \kk-\tau (\vv\ind{p}\cdot \kk)^2 \right]\tilde\phi^*(\kk)
=-\rho_0^{-1/2}\kk^2\tilde\mcV\ind{p}(\kk),
\end{equation}
leading to the solution
\begin{equation}
\tilde\phi^*(\kk)=\frac{-\rho_0^{-1/2}\kk^2\tilde\mcV\ind{p}(\kk)}{\kk^2(T+\tilde\mcV(\kk))-i\vv\ind{p}\cdot \kk-\tau (\vv\ind{p}\cdot \kk)^2 }.
\end{equation}
Finally, the density variation $\delta\rho^*(\xx)=\rho_0^{-1/2}\phi^*(\xx)$ is, in Fourier space,
\begin{equation}\label{eq:density_around_probe}
\frac{\widetilde{\delta\rho}^*(\kk)}{\rho_0}=\frac{-\kk^2\tilde V\ind{p}(\kk)}{\kk^2(T+\rho_0\tilde V(\kk))-i\vv\ind{p}\cdot \kk-\tau (\vv\ind{p}\cdot \kk)^2 }.
\end{equation}

\subsection{Drag coefficient}\label{sub:drag}

The drag force can be expressed with the density variation in Fourier space as
\begin{align}
\FF &=-\rho_0^{-1/2}\int i\kk\tilde\mcV\ind{p}(\kk)\tilde\phi^*(\kk) \frac{d\kk}{(2\pi)^d},\\
&=i\rho_0^{-1} \int\frac{\kk \kk^2\tilde\mcV\ind{p}(\kk)^2}{\kk^2(T+\tilde\mcV(\kk))-i\vv\ind{p}\cdot \kk-\tau(\vv\ind{p}\cdot \kk)^2}\frac{d\kk}{(2\pi)^d},\\
&=-\rho_0^{-1}\int \frac{\kk(\vv\ind{p}\cdot \kk) \kk^2\tilde\mcV\ind{p}(\kk)^2}{\left[\kk^2(T+\tilde\mcV(\kk))-\tau(\vv\ind{p}\cdot \kk)^2 \right]^2+(\vv\ind{p}\cdot \kk)^2}\frac{d\kk}{(2\pi)^d}.
\end{align}
Decomposing the wavevector as $\kk=(k_\parallel,\kk_\perp)$ according to the velocity $\vv\ind{p}$ allows to write the force as
\begin{equation}
\FF = -\rho_0^{-1}\vv\ind{p}\int \frac{k_\parallel^2 \kk^2\tilde\mcV\ind{p}(\kk)^2}{\left[\kk^2(T+\tilde\mcV(\kk))-\tau(v\ind{p} k_\parallel)^2 \right]^2+(v\ind{p} k_\parallel)^2}\frac{d\kk}{(2\pi)^d}.
\end{equation}
The drag coefficient $\lambda$ is defined by $\FF=-\lambda \vv\ind{p}$; it reads
\begin{equation}\label{eq:drag_coefficient}
\lambda = \rho_0\int \frac{k_\parallel^2 \kk^2\tilde V\ind{p}(\kk)^2}{\left[\kk^2(T+\rho_0\tilde V(\kk))-\tau(v\ind{p} k_\parallel)^2 \right]^2+(v\ind{p} k_\parallel)^2}\frac{d\kk}{(2\pi)^d}.
\end{equation}
This expression is our main result.
Note that this drag coefficient is induced by the bath colloids only, the total drag force also includes the drag induced by the solvent.


We are interested in the stationnary bath density around the probe (Eq.~\ref{eq:density_around_probe}) and the drag coefficient (Eq.~\ref{eq:drag_coefficient}) in two cases:
\begin{enumerate}[(i)]
\item the probe is much larger than the bath particles,
\item the probe is identical to the bath particles.
\end{enumerate}

In the case (i), the wavevectors $\kk$ that contribute to the integral in Eq.~\ref{eq:drag_coefficient} are of order $k\sim 1/a\ind{p}\ll 1$. At this scale, the Fourier transform of the bath pair potential is almost constant, $\tilde V(k)\simeq\tilde V(0)$ for $k\sim 1/a\ind{p}$.
This allows to define the ``collective'' diffusion coefficient of the bath~\cite{Sriram2010},
\begin{equation}
D\ind{bath}=T+\rho_0\tilde V(0),
\end{equation}
that sets the relaxation time of the density field $\rho(\xx,t)$.
The Péclet number is defined as
\begin{equation}
\mathrm{Pe}= \frac{a\ind{p} v\ind{p}}{D\ind{bath}};
\end{equation}
it compares the velocity of the probe to the relaxation of the density field~\cite{Sriram2010}. 

The density around the probe can then be written
\begin{equation}
\frac{\widetilde{\delta\rho}^*(\kk)}{\rho_0}=\frac{1}{D\ind{bath}}\frac{-\kk^2\tilde V\ind{p}(\kk)}{\kk^2-i \frac{\mathrm{Pe}}{a\ind{p}}k_\parallel-\tau D\ind{bath}\frac{\mathrm{Pe}^2}{a\ind{p}^2}k_\parallel^2}.
\end{equation}
To rescale the lengths to the probe size $a\ind{p}$, we introduce $\qq = a\ind{p}\kk$ and $\bar V\ind{p}(\yy)=V\ind{p}(a\ind{p}\yy)$ (so that $\tilde V\ind{p}(\kk)=a\ind{p}^d\tilde{\bar V}\ind{p}(a\ind{p}\kk)$).
With these notations, the density around the probe reads
\begin{equation}
\frac{\widetilde{\delta\rho}^*(\qq/a\ind{p})}{\rho_0}=\frac{a\ind{p}^d}{D\ind{bath}}\frac{-\qq^2\tilde{\bar V}\ind{p}(\qq)}{\qq^2-i\Pe q_\parallel - \frac{\tau}{\tau\ind{rel}}\Pe^2 q_\parallel^2},
\end{equation}
where 
\begin{equation}
\tau\ind{rel} = \frac{a\ind{p}^2}{D\ind{bath}}
\end{equation}
is the relaxation time of the field on the lengthscale of the probe.
With the appropriate normalization, this density depends only on the Péclet number and the inertial number $\tau/\tau\ind{rel}$.
This is also true for the drag coefficient, which can be written
\begin{equation}\label{eq:drag_coefficient_large_probe}
\lambda = \frac{\rho_0 a\ind{p}^d}{D\ind{bath}^2}\int \frac{q_\parallel^2\qq^2 \tilde{\bar V}\ind{p}(\qq)^2}{\left(\qq^2-\frac{\tau}{\tau\ind{rel}}\Pe^2 q_\parallel^2 \right)^2+\Pe^2 q_\parallel^2}\frac{d\qq}{(2\pi)^d}.
\end{equation}



\subsection{Numerical computation}\label{sub:numerical_computation}

The dimension is set to $d=3$ and the pair potentials are Gaussian,
\begin{align}
V(\xx) & = \epsilon \exp \left(-\frac{\xx^2}{2} \right),\\
V\ind{p}(\xx) & = \epsilon\ind{p} \exp \left(-\frac{\xx^2}{2a\ind{p}^2} \right).
\end{align}

\begin{figure*}
\begin{center}
\includegraphics[width=\linewidth]{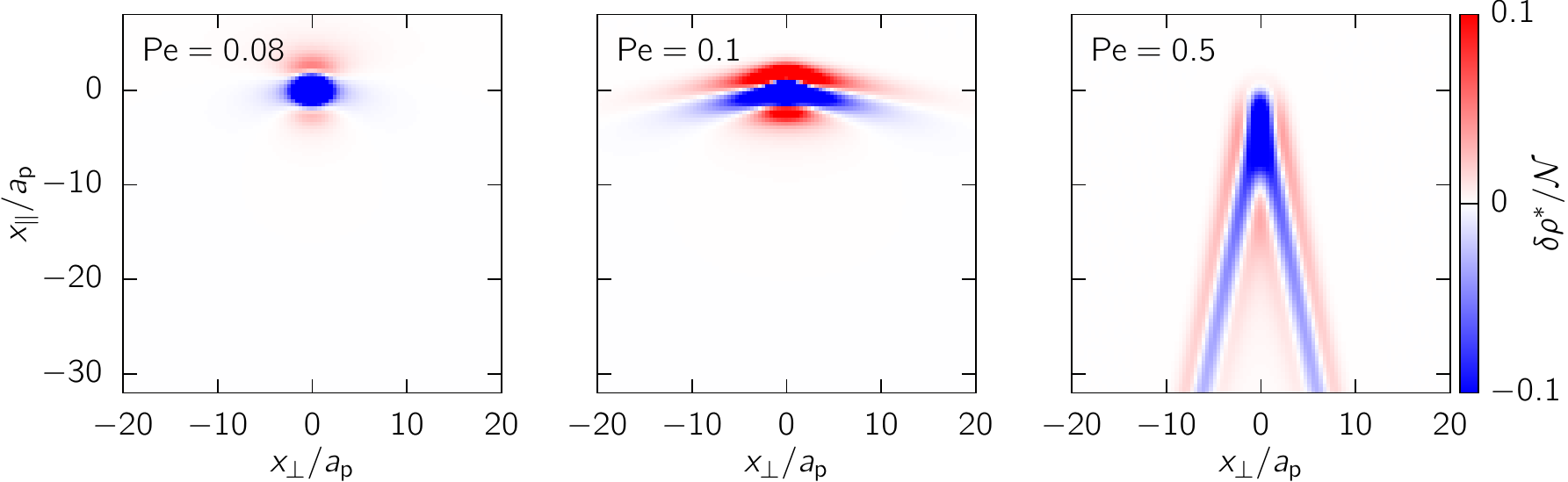}
\end{center}
\caption{(color online) Bath density variations in the reference frame of the probe, $\delta\rho^*(\xx)=\rho^*(\xx)-\rho_0$, for different values of the Péclet number, when the probe is much larger than the bath particles and the bath particles are heavy, $\tau/\tau\ind{rel}=100$. The normalization constant is $\mathcal{N}=\epsilon\ind{p}/D\ind{bath}$.}
\label{fig:bath_density_pl}
\end{figure*}

\begin{figure}
\begin{center}
\includegraphics[width=.6\linewidth]{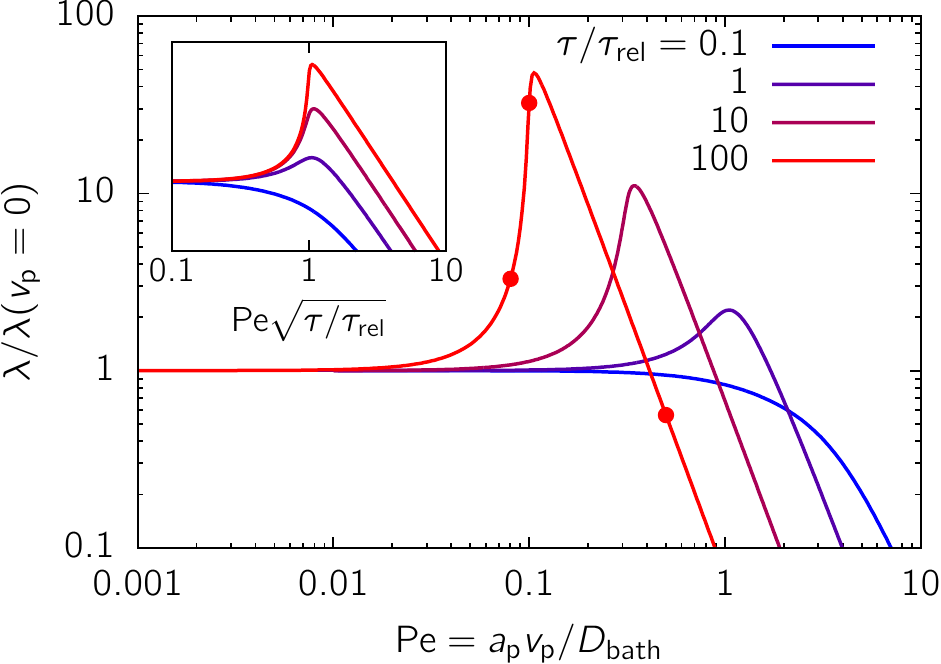}
\end{center}
\caption{(color online) Friction coefficient when the probe is much larger than the bath particles as a function of the Péclet number for different values of the inertial number $\tau /\tau\ind{rel}$, normalized by its value at small Péclet number.
The dots indicate the values used for the density profiles in Fig.~\ref{fig:bath_density_pl}.
\emph{Inset:} Normalized friction coefficient as a function of the Péclet number rescaled by $\sqrt{\tau/\tau\ind{rel}}$.}
\label{fig:v_friction_tau_pl}
\end{figure}

The bath density is plotted on Fig.~\ref{fig:bath_density_pl} for heavy bath particles ($\tau/\tau\ind{rel}=100$) at different Péclet numbers.
The drag coefficient is plotted as a function of the Péclet number on Fig.~\ref{fig:v_friction_tau_pl}, for different values of the inertial number. 
These curves exhibit shear-thickening and resemble those obtained by numerical simulations in~\cite{Kawasaki2014} (see Fig. 2a). This effect is due to inertia, as it disappears at low inertial numbers, in accordance with the conclusions of~\cite{Kawasaki2014}. 
The density profiles differ strongly from those found in~\cite{Squires2005, Carpen2005, Swan2013, Demery2014c} without inertia, or in experiments~\cite{Meyer2006,Sriram2010}. In the experiments of~\cite{Meyer2006}, one can estimate $\tau/(a_b^2/D_0)\simeq 10$ ($D_0=k\ind{B}T/\lambda$ is the thermal diffusion coefficient) so that inertia should matter; however, the bath particles are hard and the model discussed here may not apply.
Inertia allows density waves to propagate through the bath leading to the cone visible for $\mathrm{Pe}=0.1,\,0.5$ on Fig.~\ref{fig:bath_density_pl}.


\begin{figure*}
\begin{center}
\includegraphics[width=\linewidth]{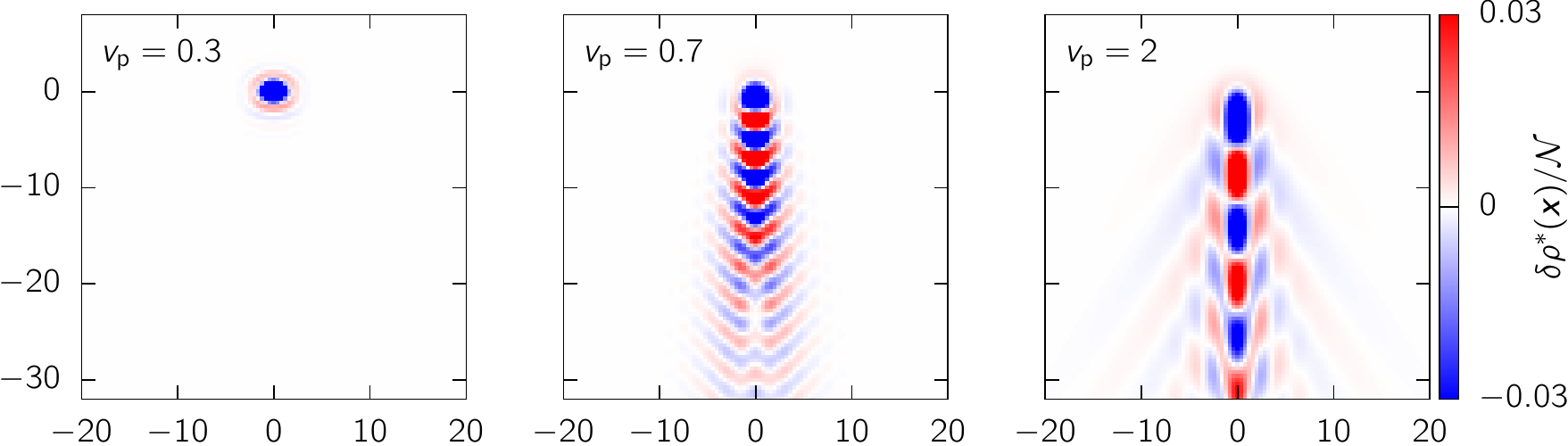}
\end{center}
\caption{(color online) Bath density variations in the reference frame of the probe, $\rho^*(x)/\rho_0-1$, for different values of the probe velocity when the bath particles are identical to the probe. The bath particles are heavy, $\tau=100$, and $\rho_0\epsilon=1$.
}
\label{fig:bath_density_same}
\end{figure*}

\begin{figure}
\begin{center}
\includegraphics[width=.6\linewidth]{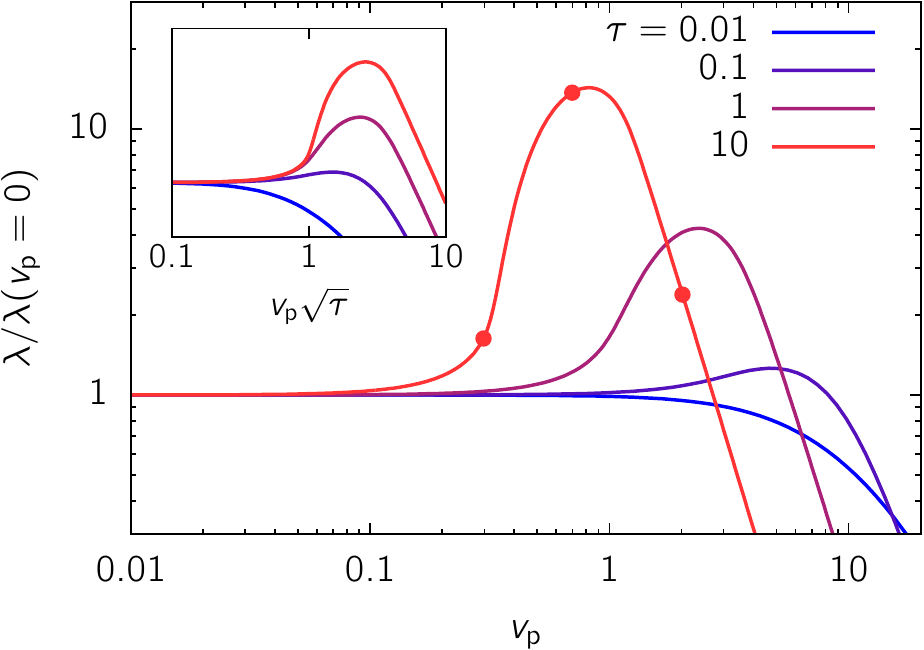}
\end{center}
\caption{(color online) Friction coefficient as a function of the probe velocity when the bath particles are identical to the probe, for different values of the inertial relaxation time $\tau$, with $\rho_0\epsilon=1$.
The friction coefficient is normalized by its value at small velocity.
The dots indicate the values used for the density profiles in Fig.~\ref{fig:bath_density_same}.
\emph{Inset:} Normalized friction coefficient as a function of the probe velocity rescaled by $\sqrt{\tau}$.}
\label{fig:v_friction_tau_samesize}
\end{figure}

\begin{figure}
\begin{center}
\includegraphics[width=.6\linewidth]{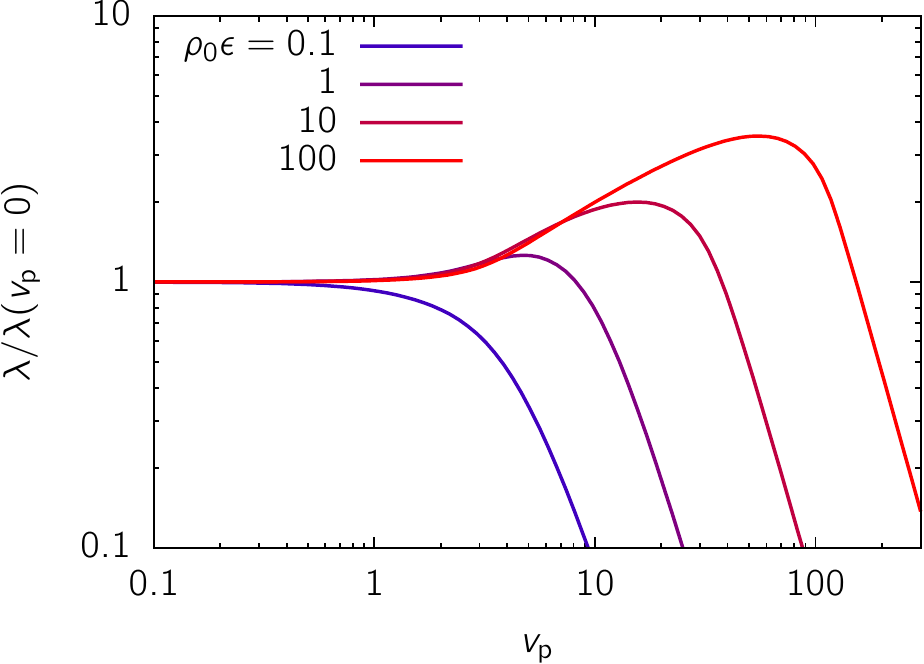}
\end{center}
\caption{(color online) Friction coefficient as a function of the probe velocity when the bath particles are identical to the probe, for different values of the interaction strength $\rho_0\epsilon$, with $\tau=0.1$. The friction coefficient is normalized by the density of the bath.}
\label{fig:v_friction_dens_samesize}
\end{figure}

In the case (ii), the dynamics of the bath depends on the mode $\kk$ and no collective diffusion coefficient can be defined. 
The bath density is plotted on Fig.~\ref{fig:bath_density_same} for $\rho_0\epsilon=1$, $\tau=100$, and different probe velocities; 
its structure is more complex than in the case (i), because the bath is dispersive at the scale of the colloids.
The drag coefficient is plotted as a function of the probe velocity on Fig.~\ref{fig:v_friction_tau_samesize} for $\rho_0\epsilon=1$ and different inertial times $\tau$, and on Fig.~\ref{fig:v_friction_dens_samesize} for $\tau=0.1$ and different interaction strengths $\rho_0\epsilon$.
A shear-thickening regime emerges at large inertial times and large interaction strengths, consistently with our finding for point-like bath particles.

\subsection{Timescales and scaling laws}\label{sub:timescales}

We show that the different observed behaviors can be rationalized by comparing the different timescales involved in our process.
We focus on the case (ii) where the probe is identical to the bath particles, $a\ind{p}=a=1$. 
Four timescales emerge in our analysis:
\begin{itemize}
\item The inertial timescale $\tau\ind{i}=m/\lambda\ind{s}=\tau$.
\item The thermal diffusion timescale $\tau\ind{th}=\lambda\ind{s} a\ind{p}^2/T=1$,
\item The timescale associated with the density relaxation due to pair interactions, $\tau\ind{pair}=\lambda\ind{s}/(\rho_0 a\ind{p}\epsilon)=1/(\rho_0\epsilon)$. The denominator, $\rho_0 a\ind{p}^3 \epsilon$, is the energy scale seen by one bath particle. 
This expression differs slightly from the one given in~\cite{Kawasaki2014} for the viscous damping timescale ($\tau_0$ in~\cite{Kawasaki2014}), where $\rho_0\simeq a\ind{p}^{-3}$ and $\tau\ind{pair}=\lambda\ind{s}a\ind{p}^2/\epsilon$.
\item The timescale associated with the motion of the probe, $\tau\ind{p}=a\ind{p}/v\ind{p}=1/v\ind{p}$. 
In macrorheology, the characteristic timescale of the forcing is set by the shear rate $\dot\gamma$: $\tau\ind{shear}=1/\dot\gamma$.
\end{itemize}

At moderate density and low temperature, the suspension undergoes a glass transition~\cite{Kawasaki2014}.
We are interested in higher densities and temperatures, where the system still behaves as a fluid~\cite{Stillinger1976,Louis2000,Ikeda2011,Berthier2010b}.
In this case, thermal diffusion and pair interactions act together to relax the density; the first two terms on the r.h.s of Eq.~\ref{eq:linearized_id} show that the two associated timescales $\tau\ind{th}$ and $\tau\ind{pair}$ are combined in one timescale $\tau\ind{rel}$ related to the relaxation of the density field:
\begin{equation}
\frac{1}{\tau\ind{rel}}=\frac{1}{\tau\ind{th}}+\frac{1}{\tau\ind{pair}}=T+\rho_0\epsilon.
\end{equation}

Without inertia, i.e. $\tau\ind{i}=0$, only two regimes are present (see Fig.~\ref{fig:v_friction_tau_samesize}, $\tau=0.1$): a Newtonian regime at small velocities where $\tau\ind{p}>\tau\ind{rel}$, and a shear-thinning regime at large probe velocities such that $\tau\ind{p}<\tau\ind{rel}$. 
The suspension shear thins because it does not have enough time to respond to the presence of the probe. This effect has already been evidenced in this framework in~\cite{Demery2014c} and for a probe in more general environments in~\cite{Demery2010}.
Shear-thinning is also observed in experiments~\cite{Meyer2006,Sriram2010}, numerical simulations~\cite{Carpen2005,Kawasaki2014} and computations~\cite{Fuchs2002,Squires2005,Gazuz2009,Gnann2011,Harrer2012,Gazuz2013}; it is due to the disruption of the static organization of the medium by the probe, which occurs when the forcing timescale becomes comparable to the relaxation timescale.

When inertia is added, expanding the denominator of the integrand in Eq.~\ref{eq:drag_coefficient} for small probe velocities leads to the following criterion: shear-thickening is present if 
\begin{equation}
\tau\ind{i}\gtrsim\tau\ind{rel}. 
\end{equation}
This is also the condition for the mode $k=\pi/a\ind{p}$ of the density to oscillate and waves to develop in the wake of the probe (see Figs.~\ref{fig:bath_density_pl} and \ref{fig:bath_density_same}).
This expansion gives the variation of the friction coefficient at small velocity as
\begin{equation}
\lambda(v\ind{p})-\lambda(v\ind{p}=0)\underset{v\ind{p}\rightarrow 0}{\sim}\left(\frac{\tau\ind{i}}{\tau\ind{rel}}-1 \right)v\ind{p}^2;
\end{equation}
this scaling law has been observed in~\cite{Kawasaki2014}, but it does not hold when the maximal drag coefficient is approached here.
It is also different from the scaling $\lambda\sim v\ind{p}$ obtained by Bagnold~\cite{Bagnold1954}, where the dissipation is dominated by the collisions between the grains.
Note that the exponent is dictated by symmetry in our model: the drag coefficient is an even function of $v\ind{p}$, so that the first correction is of order $v\ind{p}^2$.

A direct look at the same expression shows that the viscosity is maximal when
\begin{equation}\label{eq:scaling_max_viscosity}
\tau\ind{p}\simeq\sqrt{\tau\ind{i}\tau\ind{rel}},
\end{equation}
which corresponds to a resonance between the forcing and the excited mode.
This expression does not match the scaling found in~\cite{Kawasaki2014} and explained by arguments from kinetic theory, which is $\tau\ind{p}\simeq \tau\ind{i}^{3/4}\tau\ind{rel}^{1/4}$. However, Kawasaki et al. suggest that the scaling~(\ref{eq:scaling_max_viscosity}) can be obtained assuming ``soft particles and collisional dissipation''.

As the inertia $\tau$ increases, the shear-thickening becomes sharper in Figs.~\ref{fig:v_friction_tau_pl} and \ref{fig:v_friction_tau_samesize}.
Rescaling the velocity by $\sqrt{\tau}$ in Eq.~\ref{eq:drag_coefficient} (or the Péclet number by $\sqrt{\tau/\tau\ind{rel}}$ in Eq.~\ref{eq:drag_coefficient_large_probe}), it appears that the integral diverges for $\tau\ind{p}>\sqrt{\tau\ind{i}\tau\ind{rel}}$ when $\tau\rightarrow\infty$.
This divergence is shown in the inset of Figs.~\ref{fig:v_friction_tau_pl} and \ref{fig:v_friction_tau_samesize}; it shows that the shear-thickening becomes discontinuous in this limit.

At high velocities such that $\tau\ind{p}\ll\sqrt{\tau\ind{i}\tau\ind{rel}}$, the drag coefficient induced by the bath particles decays as $\sim v\ind{p}^{-4}$.
The total drag force thus reduces to the drag force induced by the solvent.

\subsection{Constant velocity vs. constant force}\label{sub:cv_cf}

As pointed out above, the drag coefficient depends on the mode of driving: constant force or constant velocity~\cite{Swan2013,Puertas2014}.
We show this effect in the limit of zero velocity, or zero force, and without inertia ($\tau=0$). The drag coefficient (\ref{eq:drag_coefficient}) reduces to
\begin{equation}\label{eq:drag_coeff_cv}
\lambda_\mathrm{c.v.} = \frac{\rho_0}{d}\int \frac{\tilde V(\kk)^2}{[T+\rho_0\tilde V(\kk)]^2}\frac{d\kk}{(2\pi)^d}.
\end{equation}
The drag coefficient at constant force has been computed in~\cite{Demery2014c} (Eq.~79); at zero force, it is
\begin{equation}\label{eq:drag_coeff_cf}
\lambda_\mathrm{c.f.} = \frac{\rho_0}{d}\int \frac{\tilde V(\kk)^2}{[T+\rho_0\tilde V(\kk)] [2T+\rho_0\tilde V(\kk)]}\frac{d\kk}{(2\pi)^d}.
\end{equation}

As found in~\cite{Swan2013}, the drag coefficient is smaller when measured at constant force, because of the factor $2$ in front of the temperature in the denominator of Eq.~\ref{eq:drag_coeff_cf}, which accounts for the diffusion of the probe~\cite{Demery2014c}.
Interestingly, in the ``dilute'' limit where $\rho_0\tilde V(\kk)\ll T$, we recover the relation $\lambda\ind{c.v.}=2\lambda\ind{c.f.}$ found in~\cite{Swan2013}.

\section{Conclusion}\label{}

We considered a suspension of colloids with inertia and showed that the drag force on a probe pulled at constant velocity can be computed analytically in the limit of a dense suspension and soft colloids.
The rheology of the suspension can be deduced: shear-thickening is observed if the inertia is large enough for density waves to propagate through the solution, in agreement with~\cite{Kawasaki2014}.
However, some quantitative differences arise between these numerical simulations and our computations in the scalings of the viscosity before the maximum and the position of the maximum. 
Numerical simulations closer to the regime that we studied, which are beyond the scope of this paper, would help to bridge this gap.

In~\cite{Demery2014c}, we shown that the ideal limit of a very dense and soft suspension reproduces the tracer diffusion coefficient computed in the dilute limit~\cite{Dean2004} and exhibits force-induced diffusion, which has also been observed for hard particles~\cite{Zia2010,Harrer2012,Benichou2013d}.
Adding inertia, we showed here that it reproduces the basic features of inertia-induced shear-thickening reported for numerical simulations for hard particles~\cite{Kawasaki2014}.
These examples show that albeit the assumptions made in the computation are restrictive, this mean-field model can shed light on generic rheological properties of colloidal suspensions. 
At this stage, when applied to systems that are not in the theoretical range of validity, the outcome of this model cannot be considered as quantitatively accurate.

\section*{Acknowledgments}\label{}

I thank L. Berthier, V. Lecomte, A. Lefèvre and M. Benzaquen for their constructive comments and our stimulating discussinos.
I acknowledge financial support by the KECK foundation Award 37086.



\begin{thebibliography}{10}

\bibitem{Mewis2011}
Jan Mewis and Norman~J. Wagner.
\newblock {\em {Colloidal suspension rheology}}.
\newblock {Cambridge University Press}, 2011.

\bibitem{Coussot2005}
Philippe Coussot.
\newblock {\em {Rheometry of pastes, suspensions, and granular materials:
  applications in industry and environment}}.
\newblock {John Wiley \& Sons}, 2005.

\bibitem{Ikeda2012}
Atsushi Ikeda, Ludovic Berthier, and Peter Sollich.
\newblock {Unified study of glass and jamming rheology in soft particle
  systems}.
\newblock {\em {Phys. Rev. Lett.}}, 109(1):018301, {Jul} 2012.

\bibitem{Laun1984}
Hans~Martin Laun.
\newblock {Rheological properties of aqueous polymer dispersions}.
\newblock {\em {Die Angewandte Makromolekulare Chemie}}, 123(1):335--359, 1984.

\bibitem{Wagner2009}
Norman~J. Wagner and John~F. Brady.
\newblock {Shear thickening in colloidal dispersions}.
\newblock {\em {Physics Today}}, 62(10):27--32, 2009.

\bibitem{Bagnold1954}
R.~A. Bagnold.
\newblock {Experiments on a Gravity-Free Dispersion of Large Solid Spheres in a
  Newtonian Fluid under Shear}.
\newblock {\em {Proceedings of the Royal Society of London A: Mathematical,
  Physical and Engineering Sciences}}, 225(1160):49--63, 1954.

\bibitem{Boyer2011b}
Fran{\c{c}}ois Boyer, {\'{E}}lisabeth Guazzelli, and Olivier Pouliquen.
\newblock {Unifying Suspension and Granular Rheology}.
\newblock {\em {Phys. Rev. Lett.}}, 107(18):188301, {Oct} 2011.

\bibitem{Waigh2005}
T.~A. Waigh.
\newblock {Microrheology of complex fluids}.
\newblock {\em {Reports on Progress in Physics}}, 68(3):685, 2005.

\bibitem{Squires2010}
Todd~M. Squires and Thomas~G. Mason.
\newblock {Fluid Mechanics of Microrheology}.
\newblock {\em {Annual Review of Fluid Mechanics}}, 42(1):413--438, 2010.

\bibitem{Habdas2004}
P.~Habdas, D.~Schaar, A.~C. Levitt, and E.~R. Weeks.
\newblock {Forced motion of a probe particle near the colloidal glass
  transition}.
\newblock {\em {EPL (Europhysics Letters)}}, 67(3):477, 2004.

\bibitem{Wilson2011b}
Laurence~G. Wilson and Wilson C.~K. Poon.
\newblock {Small-world rheology: an introduction to probe-based active
  microrheology}.
\newblock {\em {Phys. Chem. Chem. Phys.}}, 13(22):10617--10630, 2011.

\bibitem{Meyer2006}
Alexander Meyer, Andrew Marshall, Brian~G. Bush, and Eric~M. Furst.
\newblock {Laser tweezer microrheology of a colloidal suspension}.
\newblock {\em {Journal of Rheology}}, 50(1):77--92, 2006.

\bibitem{Sriram2010}
Indira Sriram, Alexander Meyer, and Eric~M. Furst.
\newblock {Active microrheology of a colloidal suspension in the direct
  collision limit}.
\newblock {\em {Physics of Fluids (1994-present)}}, 22(6), 2010.

\bibitem{Puertas2014}
A.~M. Puertas and T.~Voigtmann.
\newblock {Microrheology of colloidal systems}.
\newblock {\em {Journal of Physics: Condensed Matter}}, 26(24):243101, 2014.

\bibitem{Kawasaki2014}
Takeshi Kawasaki, Atsushi Ikeda, and Ludovic Berthier.
\newblock {Thinning or thickening?: Multiple rheological regimes in dense
  suspensions of soft particles}.
\newblock {\em {EPL}}, 107(2):28009, 2014.

\bibitem{Trulsson2014}
M.~Trulsson, M.~Bouzid, J.~Kurchan, E.~Clement, P.~Claudin, and B.~Andreotti.
\newblock {Athermal analogue of sheared colloidal suspensions}.
\newblock {\em {ArXiv e-prints}}, {nov} 2014.

\bibitem{Trulsson2012}
Martin Trulsson, Bruno Andreotti, and Philippe Claudin.
\newblock {Transition from the Viscous to Inertial Regime in Dense
  Suspensions}.
\newblock {\em {Phys. Rev. Lett.}}, 109(11):118305, {Sep} 2012.

\bibitem{Carpen2005}
Ileana~C. Carpen and John~F. Brady.
\newblock {Microrheology of colloidal dispersions by Brownian dynamics
  simulations}.
\newblock {\em {Journal of Rheology (1978-present)}}, 49(6):1483--1502, 2005.

\bibitem{Winter2012}
D.~Winter, J.~Horbach, P.~Virnau, and K.~Binder.
\newblock {Active Nonlinear Microrheology in a Glass-Forming Yukawa Fluid}.
\newblock {\em {Phys. Rev. Lett.}}, 108(2):028303, {Jan} 2012.

\bibitem{Fuchs2002}
Matthias Fuchs and Michael~E. Cates.
\newblock {Theory of Nonlinear Rheology and Yielding of Dense Colloidal
  Suspensions}.
\newblock {\em {Phys. Rev. Lett.}}, 89(24):248304, {Nov} 2002.

\bibitem{Squires2005}
Todd~M. Squires and John~F. Brady.
\newblock {A simple paradigm for active and nonlinear microrheology}.
\newblock {\em {Physics of Fluids (1994-present)}}, 17(7):073101, 2005.

\bibitem{Khair2006}
Aditya~S. Khair and John~F. Brady.
\newblock {Single particle motion in colloidal dispersions: a simple model for
  active and nonlinear microrheology}.
\newblock {\em {Journal of Fluid Mechanics}}, 557:73--117, {6} 2006.

\bibitem{Harrer2012}
Ch~J. Harrer, D.~Winter, J.~Horbach, M.~Fuchs, and Th~Voigtmann.
\newblock {Force-induced diffusion in microrheology}.
\newblock {\em {Journal of Physics: Condensed Matter}}, 24(46):464105, 2012.

\bibitem{Gazuz2009}
I.~Gazuz, A.~M. Puertas, Th. Voigtmann, and M.~Fuchs.
\newblock {Active and Nonlinear Microrheology in Dense Colloidal Suspensions}.
\newblock {\em {Phys. Rev. Lett.}}, 102(24):248302, {Jun} 2009.

\bibitem{Gnann2011}
Manuel~Victor Gnann, Igor Gazuz, Antonio~Manuel Puertas, Matthias Fuchs, and
  Th~Voigtmann.
\newblock {Schematic models for active nonlinear microrheology}.
\newblock {\em {Soft Matter}}, 7(4):1390--1396, 2011.

\bibitem{Gazuz2013}
I.~Gazuz and M.~Fuchs.
\newblock {Nonlinear microrheology of dense colloidal suspensions: A
  mode-coupling theory}.
\newblock {\em {Phys. Rev. E}}, 87(3):032304, {Mar} 2013.

\bibitem{Demery2014c}
Vincent D{\'{e}}mery, Olivier B{\'{e}}nichou, and Hugo Jacquin.
\newblock {Generalized Langevin equations for a driven tracer in dense soft
  colloids: construction and applications}.
\newblock {\em {New Journal of Physics}}, 16(5):053032, 2014.

\bibitem{Brown2014}
Eric Brown and Heinrich~M. Jaeger.
\newblock {Shear thickening in concentrated suspensions: phenomenology,
  mechanisms and relations to jamming}.
\newblock {\em {Reports on Progress in Physics}}, 77(4):046602, 2014.

\bibitem{Brown2009}
Eric Brown and Heinrich~M. Jaeger.
\newblock {Dynamic Jamming Point for Shear Thickening Suspensions}.
\newblock {\em {Phys. Rev. Lett.}}, 103(8):086001, {Aug} 2009.

\bibitem{Wyart2014}
M.~Wyart and .~E. Cates, M.~\.
\newblock {Discontinuous Shear Thickening without Inertia in Dense Non-Brownian
  Suspensions}.
\newblock {\em {Phys. Rev. Lett.}}, 112(9):098302, {Mar} 2014.

\bibitem{Fall2015}
A.~Fall, F.~Bertrand, D.~Hautemayou, C.~Mezi{\`{e}}re, P.~Moucheront,
  A.~Lema{\^{i}}tre, and G.~Ovarlez.
\newblock {Macroscopic Discontinuous Shear Thickening versus Local Shear
  Jamming in Cornstarch}.
\newblock {\em {Phys. Rev. Lett.}}, 114(9):098301, {Mar} 2015.

\bibitem{Melrose2004}
John~R. Melrose and Robin~C. Ball.
\newblock {{``}Contact networks{''} in continuously shear thickening colloids}.
\newblock {\em {Journal of Rheology (1978-present)}}, 48(5):961--978, 2004.

\bibitem{Bergenholtz2002}
J.~Bergenholtz, J.~F. Brady, and M.~Vicic.
\newblock {The non-Newtonian rheology of dilute colloidal suspensions}.
\newblock {\em {Journal of Fluid Mechanics}}, 456:239--275, {4} 2002.

\bibitem{Mewis2001}
Jan Mewis and Gary Biebaut.
\newblock {Shear thickening in steady and superposition flows effect of
  particle interaction forces}.
\newblock {\em {Journal of Rheology (1978-present)}}, 45(3):799--813, 2001.

\bibitem{Swan2013}
James~W. Swan and Roseanna~N. Zia.
\newblock {Active microrheology: Fixed-velocity versus fixed-force}.
\newblock {\em {Physics of Fluids (1994-present)}}, 25(8), 2013.

\bibitem{Nakamura2009}
Takenobu Nakamura and Akira Yoshimori.
\newblock {Derivation of the nonlinear fluctuating hydrodynamic equation from
  the underdamped Langevin equation}.
\newblock {\em {Journal of Physics A: Mathematical and Theoretical}},
  42(6):065001, 2009.

\bibitem{Das2013}
Shankar~P. Das and Akira Yoshimori.
\newblock {Coarse-grained forms for equations describing the microscopic motion
  of particles in a fluid}.
\newblock {\em {Phys. Rev. E}}, 88(4):043008, {Oct} 2013.

\bibitem{Hansen2006}
Jean-Pierre Hansen and I.~R. McDonald.
\newblock {\em {Theory of Simple Liquids}}.
\newblock {London: Academic Press}, {3rd} edition, {apr} 2006.

\bibitem{Demery2010}
Vincent D{\'{e}}mery and David~S. Dean.
\newblock {Drag Forces in Classical Fields}.
\newblock {\em {Phys. Rev. Lett.}}, 104(8):080601, {Feb} 2010.

\bibitem{Stillinger1976}
Frank~H. Stillinger.
\newblock {Phase transitions in the Gaussian core system}.
\newblock {\em {The Journal of Chemical Physics}}, 65(10):3968--3974, 1976.

\bibitem{Louis2000}
A.~A. Louis, P.~G. Bolhuis, and J.~P. Hansen.
\newblock {Mean-field fluid behavior of the Gaussian core model}.
\newblock {\em {Phys. Rev. E}}, 62(6):7961--7972, {Dec} 2000.

\bibitem{Ikeda2011}
Atsushi Ikeda and Kunimasa Miyazaki.
\newblock {Thermodynamic and structural properties of the high density Gaussian
  core model}.
\newblock {\em {The Journal of Chemical Physics}}, 135(2):024901, 2011.

\bibitem{Berthier2010b}
Ludovic Berthier, Angel~J. Moreno, and Grzegorz Szamel.
\newblock {Increasing the density melts ultrasoft colloidal glasses}.
\newblock {\em {Phys. Rev. E}}, 82(6):060501, {Dec} 2010.

\bibitem{Dean2004}
David~S. Dean and Alexandre Lef{\`{e}}vre.
\newblock {Self-diffusion in a system of interacting Langevin particles}.
\newblock {\em {Physical Review E}}, 69:061111, 2004.

\bibitem{Zia2010}
Roseanna~N. Zia and John~F. Brady.
\newblock {Single-particle motion in colloids: force-induced diffusion}.
\newblock {\em {Journal of Fluid Mechanics}}, 658:188--210, {9} 2010.

\bibitem{Benichou2013d}
O.~B{\'{e}}nichou, P.~Illien, G.~Oshanin, and R.~Voituriez.
\newblock {Fluctuations and correlations of a driven tracer in a hard-core
  lattice gas}.
\newblock {\em {Phys. Rev. E}}, 87(3):032164, {Mar} 2013.

\end{thebibliography}

\end{document}